A Submission to the NSF MPSAC ngGW Subcommittee

# COSMIC EXPLORER

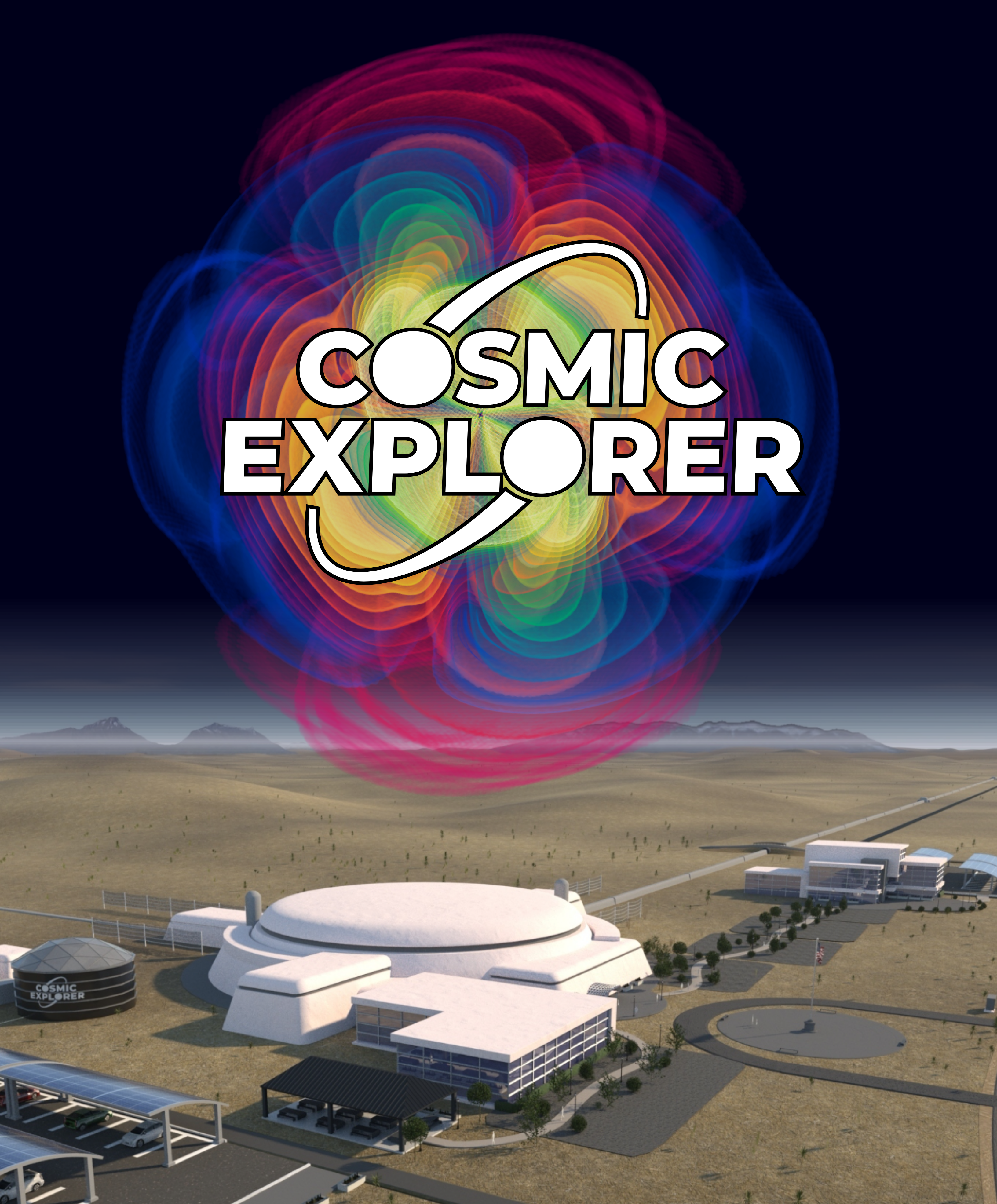

## Executive Summary

Gravitational-wave astronomy has revolutionized humanity's view of the universe, a revolution driven by observations that no other field can make. This white paper describes an observatory that builds on decades of investment by the National Science Foundation and that will drive discovery for decades to come: **Cosmic Explorer**. Major discoveries in astronomy are driven by three related improvements: better sensitivity, higher precision, and opening new observational windows. Cosmic Explorer promises all three and will deliver an order-of-magnitude greater sensitivity than LIGO. Cosmic Explorer will push the gravitational-wave frontier to almost the edge of the observable universe using technologies that have been proven by LIGO during its development.

With the unprecedented sensitivity that only a new facility can deliver, Cosmic Explorer will make discoveries that cannot yet be anticipated, especially since gravitational waves are both synergistic with electromagnetic observations *and* can reach into regions of the universe that electromagnetic observations cannot explore. With Cosmic Explorer, scientists can use the universe as a laboratory to test the laws of physics and study the nature of matter. Cosmic Explorer allows the United States to continue its leading role in gravitational-wave science and the international network of next-generation observatories. With its extraordinary discovery potential, Cosmic Explorer will deliver revolutionary observations across astronomy, physics, and cosmology including:

**Black Holes and Neutron Stars Throughout Cosmic Time.** Understanding the birth and growth of the first black holes is an important unsolved problem in astrophysics. Cosmic Explorer will detect binaries containing the first black holes, providing a view of the Cosmic Dawn complementary to that of the James Webb Space Telescope. By observing millions of compact-object mergers across the history of the universe, Cosmic Explorer will map the populations of neutron stars and black holes across time, bringing new insights into birth, life, and death of massive stars.

**Multi-Messenger Astrophysics and Dynamics of Dense Matter.** Through exquisite measurements of the interior structure of thousands of neutron stars in mergers, Cosmic Explorer will probe the nature of high-density matter and the strong nuclear force, revealing the nuclear equation of state and its phase transitions in unprecedented detail. The hot, dense remnants of neutron-star mergers will map unexplored regions of the quantum chromodynamics phase space. A plethora of multi-messenger observations will help scientists understand the production of the chemical elements that are the building blocks of the universe, and reveal the physics powering short gamma-ray bursts.

**New Probes of Extreme Astrophysics.** Neutron stars, black holes, and supernovae are expected to produce gravitational-wave signals that have not yet been observed by LIGO or Virgo — and may lie beyond their reach. LIGO and Virgo are already detecting signals from merging systems that we do not fully understand; Cosmic Explorer will reveal the nature of these mysterious sources. By detecting new types of sources, Cosmic Explorer can explore extreme astrophysical phenomena.

**Fundamental Physics and Precision Cosmology.** Cosmic Explorer's extraordinary sensitivity will allow observations of loud and rare events that will explore the nature of spacetime, point the way to a quantum theory of gravity, and uncover the unusual and unexpected. Cosmic Explorer will make on order-of-magnitude leap in precision measurements of the cosmic expansion history.

**Dark Matter and the Early Universe.** Cosmic Explorer will probe the nature of dark matter through its possible signature in mergers or black-hole superradiance in a complementary way to searches at high-energy colliders and direct-detection experiments. Cosmic Explorer provides an opportunity to observe the early universe via a cosmological stochastic gravitational-wave background.

Cosmic Explorer will be realized using a technology demonstrated by LIGO: the dual-recycled Fabry–Perot Michelson interferometer. Cosmic Explorer's factor of ten increase in sensitivity comes primarily from scaling up the detector's length from 4 to 40 km. This increases the amplitude of the observed signals with effectively no increase in the detector noise. Targeted improvements to specific detector technologies will be developed over the coming decade that will enable Cosmic Explorer to fully realize its sensitivity. These technologies can be tested in the existing LIGO facilities, ensuring technical readiness, by creating a pathfinder observatory known as LIGO A$^\sharp$ that will allow us to explore out to much greater distances than before. Figure 1 shows the tremendous astrophysical reach of Cosmic Explorer and compares this with LIGO A+ and A$^\sharp$. Cosmic Explorer's facilities will be long-lived, allowing for detector upgrades with technologies yet to be discovered.

The National Science Foundation-funded *Cosmic Explorer Horizon Study* [1] presents a concept for Cosmic Explorer consisting of a 40 km observatory and a 20 km observatory, both located in the continental United States, with a total estimated cost of $1.6B (2021 USD). While initial studies indicate that many locations could accommodate facilities of this scale, site evaluation and identification will require broader and deeper studies that consider the environmental, cultural, socio-economic, and political impacts. Building partnerships with the local and Indigenous communities is an essential part of Cosmic Explorer's mission and will be critical for ensuring that Cosmic Explorer respects, supports, and engages with its host communities. Following the successful example of the LIGO Livingston Observatory in Louisiana, Cosmic Explorer presents an opportunity to broaden participation and build research competitiveness in states that have historically been awarded less National Science Foundation support.

Cosmic Explorer is an opportunity for tremendous investment in the United States' scientific workforce and new industrial partnerships needed to realize the observatories. The team leading Cosmic Explorer draws many of its members from three Hispanic-Serving Institutions (California State University Fullerton, Texas Tech University, and University of Arizona) and three Emerging Research Institutions (California State University Fullerton, Syracuse University, and Texas Tech University). Cosmic Explorer will build on its record of supporting people historically excluded from STEM careers and will help create a diverse and inclusive twenty-first century workforce.

Cosmic Explorer, as envisioned in the Horizon Study, will achieve the majority of its science goals without the involvement of other gravitational-wave detectors. This white paper explores the synergies and dependencies on other facilities, including LIGO A+ and the potential A$^\sharp$ upgrade. Two Cosmic Explorer observatories operating together with a single 4 km LIGO detector at A$^\sharp$ sensitivity will allow the United States to independently achieve Cosmic Explorer's full range of science goals. However, Cosmic Explorer's scientific output will be greatly enhanced by operating as part of an international, multi-messenger network of gravitational-wave observatories, astro-particle detectors, and telescopes across the electromagnetic spectrum. Cosmic Explorer will bring an unprecedented view of the universe to this network.

With sustained funding and no major delays, Cosmic Explorer's first observing runs will take place in the mid-2030s — by which time any envisioned upgrades to LIGO will have approached the limit of a 4 km facility. Cosmic Explorer's operations phase will encompass observing, disseminating data, and generating astronomical alerts. Cosmic Explorer will generate a unique, rich, and deep view of the universe over its lifetime. With foundations laid by decades of National Science Foundation investment and the work of a large community of scientists, Cosmic Explorer is poised to propel another revolution in our understanding of the universe.

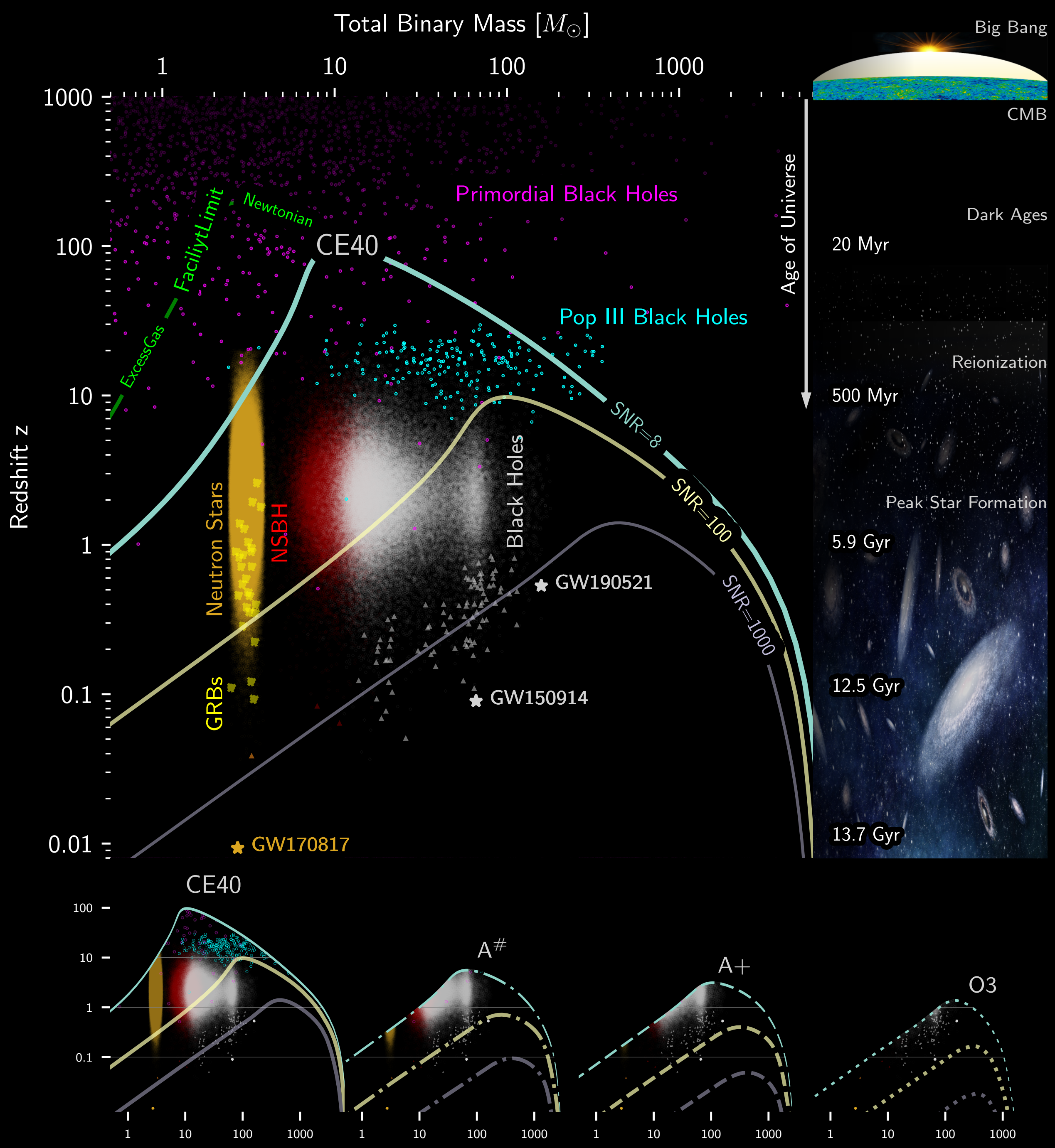

**Figure 1:** The reach of the Cosmic Explorer 40 km observatory for compact binary mergers as a function of total binary mass and redshift at various signal-to-noise ratio (SNR) thresholds. Cosmic Explorer will push the cosmic horizon to the boundary of the population of binary neutron stars (gold), neutron star – black holes (NSBH) (red) and binary black hole mergers (white) (§1.1). The order of magnitude improvement in sensitivity enables observation of new populations, including mergers from Population III black holes (cyan), and speculative primordial black holes (magenta) [2–5]. A sample of observed short gamma-ray burst (GRB) redshifts [6] is shown (yellow, with masses drawn from the BNS population). SNR > 100 signals (below yellow curve) will enable precision astrophysics (§§ 1.2 and 1.4). GW170817, GW150914, and GW190521 (stars) are highlighted along with the population of observed compact-object binaries (small triangles) [7, 8]. The facility limit (green, see §2) is shown with limiting noise sources; upgrades beyond the initial concept may approach this limit. A comparison to A$^\sharp$, A+, and O3 is shown at the bottom.

*Cosmic Explorer: A Submission to the NSF MPSAC ngGW Subcommittee*
Technical Report CE–P2300018–v3
June 2023

# Contents

In this white paper, pages that are considered part of the twenty-page submission are numbered with Arabic numerals. Pages that are not part of the page count are left blank (cover page and table of contents) or labeled with Roman numerals (acronyms, author list, and references).

# 1. Key Science Objectives

The National Academies' Decadal Survey on Astronomy and Astrophysics 2020 (Astro2020) has highlighted Cosmic Explorer (CE) as "*...central to achieving the science vision laid out in the survey's roadmap*" [9] and the Gravitational-Wave International Committee (GWIC) Science Book [10] states that "*A global next-generation gravitational wave observatory will propel the field of astrophysics and all foundational science research forward.*" CE, taking advantage of completely new facilities and with a strain sensitivity ten times that of the Laser Interferometer Gravitational Wave Observatory (LIGO) A+ upgrade (Fig. 1), will open new discovery space across five major scientific areas—CE's key science objectives (Fig. 2). Building on the Astro2020 and GWIC reports, and with input from the scientific community, the Cosmic Explorer Horizon Study (CEHS) described three central targets for CE's science. Since the publication of the CEHS, engagement with the scientific community has continued through the Dawn Workshop [11], the Snowmass2021 Community Planning Exercise [12–14], and a call to the CE Consortium for science letters that expand and update the CEHS' science goals [CESL1–CESL19].

In this Section, we describe CE's key science objectives, summarizing the work of the GWIC Science Book, the Horizon Study, and subsequent community engagement. To illustrate the transformative scale of CE's sensitivity improvement, we contrast it with LIGO A+ and A$^\sharp$, the latter being representative of the anticipated facility limit of LIGO due to its 4 km arm length [15–17]. In §3, we discuss the impact of various observatory[1] configurations across these science objectives.

## 1.1. Black Holes and Neutron Stars Throughout Cosmic Time

Merging compact-object binaries containing neutron stars (NSs) and black holes (BHs) are the engine of the most energetic events in the universe [18–20]. Gravitational wave (GW) observations will continue to revolutionize our understanding of where and when NSs and BHs formed, shed light on the life and death of massive stars, and explain the evolution and properties of their host galaxies. CE will explore the variety of astrophysical formation scenarios proposed for compact binaries through precise measurement of the population's properties, including the masses and spins of the compact objects and eccentricity of the binary systems. CE will map the merger rate histories as a function of redshift, distinguishing between the various models proposed [3, 4, 21–33]. There is already some evidence that the observed variety of source properties is likely the result of multiple astrophysical formation channels [34–37]. The peak of the mass function, and its variation with redshift, contains crucial clues about binary evolution and the final stages of the life of massive stars [CESL1, 38–43]. Fig. 1 compares and contrasts the reach of LIGO A+, A$^\sharp$, and CE. In the A+ configuration, LIGO will detect only a small fraction of the 10–10 $M_\odot$ binary black holes (BBHs) (near the peak of the local mass function) [44–46], while CE will detect the majority of such mergers in the universe, fully mapping the population.

With networks including two CE observatories, precise measurements of the binary merger rate, mass, and spin distribution across a large range of redshifts will probe the impact of stellar environments on compact binary yields and merger delay times, and ultimately untangle their formation channels and physics [5, 47–54] (Fig. 2, top; Fig. 4, upper left). For many thousands of binaries every year, CE will precisely measure the spins and masses of the individual BHs [55], shedding light on binary mass transfer stability and efficiency [CESL9, 56–59], mass ratio reversal [60–62], stellar winds and mass loss [63, 64], massive stars' sizes, and more. CE may be the only way to set these constraints at high redshifts [CESL8]. Its sensitivity below 20 Hz will enhance constraints on binary formation pathways via orbital eccentricity measurements [CESL19, 65–67].

[1] In the literature, the words "observatory", "detector" and "interferometer" are often used interchangeably. Herein an *observatory* is made up of a single *facility* and the gravitational wave (GW) *detector* it hosts. The *detector* has many sub-systems, at the center of which is an optical *interferometer*.

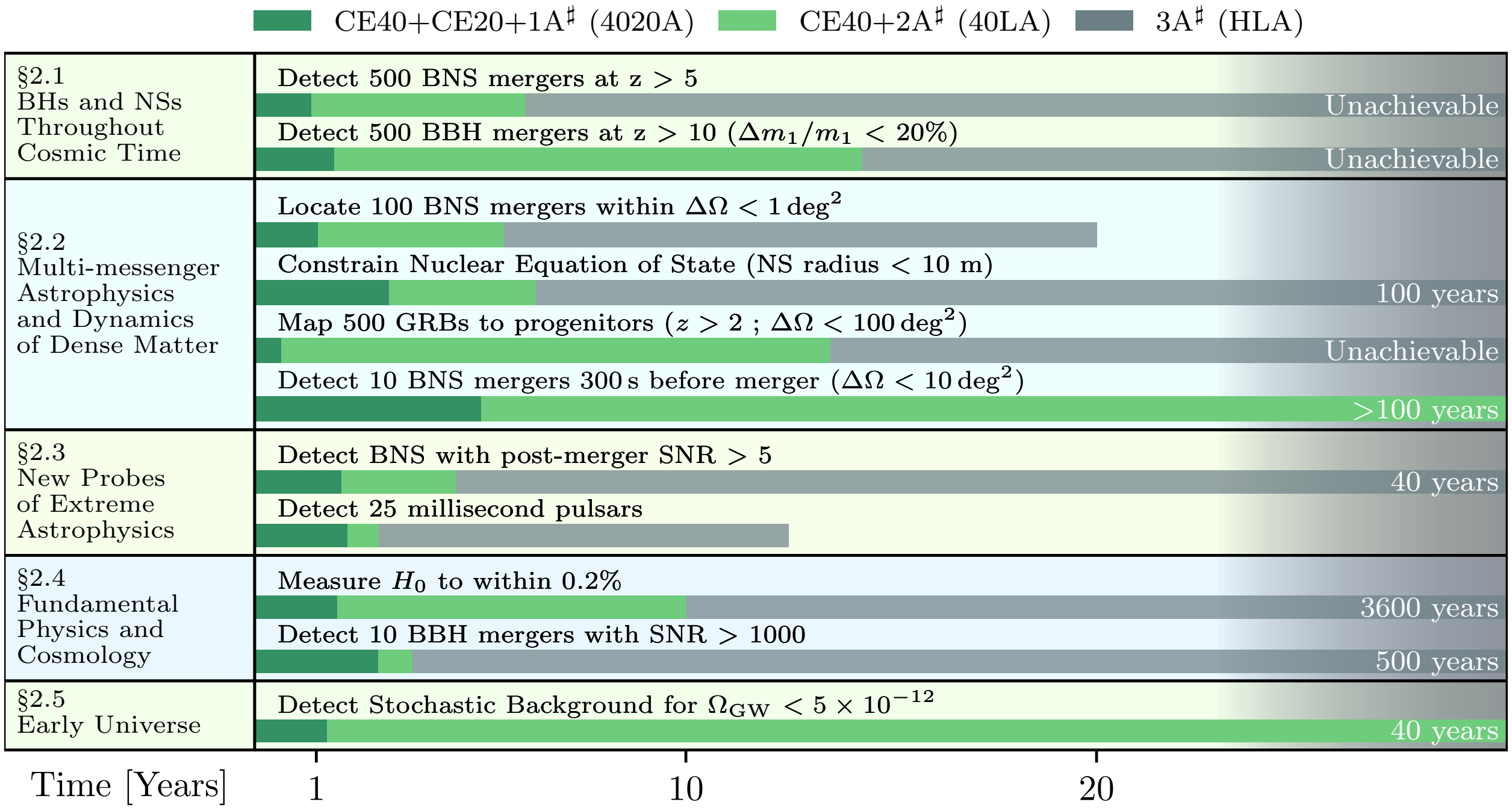

**Figure 2:** Time needed to achieve key science goals (§1). A 40 km CE in a network either with a 20 km CE and an A$^\sharp$ observatory (dark green; see 4020A in §3.2) or two A$^\sharp$ observatories (light green; see 40LA in §3.2) are compared to a network of three A$^\sharp$ observatories (gray; see HLA in §3.2). All metrics use the broadband configuration for the 20 km CE except the post-merger SNR which uses its kilohertz-focused mode (see Fig. 3). Times that exceed the plot range are given on the right. CE in its reference design, operating with one 4 km LIGO detector at A$^\sharp$ sensitivity, will ensure that the US can self-sufficiently achieve the key science goals described in §1.

Constraining the BH mass function above 50 $M_\odot$ will allow for a better understanding of the pair instability supernova mass gap (and of the nuclear physics processes that lead to it) [CESL10, 68]; the rate of hierarchical mergers [69, 70]; and intermediate mass black hole binaries (IMBHs) [CESL3, 71]. A$^\sharp$ will observe a 100–100 $M_\odot$ IMBH merger up to $z \sim 4$, missing the critical BBHs at higher redshifts (Fig. 1). CE, on the other hand, can probe formation and merger of BHs created by Pop III stars and other possible high-redshift channels, e.g., primordial black holes (PBHs) created during the inflationary epoch of the universe [3, 72, 73] (Fig. 4, upper left). Little is known about Pop III stars (see [74] for a review), and while the James Webb Space Telescope (JWST) might provide some information in the next few years [75], it will not directly resolve individual Pop III stars. The mass functions of both PBHs and remnants of Pop III stars are uncertain; but they might be the seeds that formed the supermassive BHs found at the centers of most galaxies [71, 76–79]; CE will thus explore one of the most pressing open questions in galaxy and structure formation. Since a network is required to accurately measure the mass of a distant source (§3), two CE observatories are required to map the mass distribution of BBHs at $z \gtrsim 5$, where LIGO A$^\sharp$ would have signal-to-noise ratio (SNR) less than 5 for a 10–10 $M_\odot$ BBH [47, 52, 53, 80].

Similar considerations apply to binary neutron stars (BNSs). CE will map the properties of BNSs across cosmic history and galactic environments (Fig. 1). By establishing the rate and distribution of BNS mergers out to cosmological distances, CE may also measure the time delay distribution between formation and merger [49], and thereby infer the history of chemical evolution in the universe beyond the reach of multi-messenger astronomy [54] (§1.2). The two CE observatories are key to precisely measuring masses, distances and sky positions of thousands of BNSs per year (Fig. 4, upper left).

## 1.2. Multi-Messenger Astrophysics and Dynamics of Dense Matter

The first BNS merger detected in GWs and across the electromagnetic spectrum (GW170817; [18, 81] and references therein) is a spectacular example of a discovery that impacted a rich variety of fields, ranging from nuclear and fundamental physics (e.g., [82–89]), to relativistic astrophysics (e.g., [81, 90–96]) and cosmology (e.g., [97–99]). With an order of magnitude in sensitivity improvement relative to A+, CE will extend the reach of multi-messenger astronomy to the high-redshift universe, unveiling BNS mergers around and beyond the peak of star formation, or $z \sim 2$ (which is inaccessible with 4 km observatories even at $A^\sharp$ sensitivity; Fig. 2), making it possible to unveil the progenitors of short gamma-ray bursts (GRBs) (Fig. 1) [100–102]. BNSs up to $z \sim 2$ can be localized to better than 10 $\mathrm{deg}^2$ in the sky with an international network containing the two CE observatories [103] (Fig. 4, upper right). The two CE observatories will detect hundreds of BNS mergers per year with SNR > 100 (Fig. 4, top left) determining their properties with unprecedented precision [104–106]. Measurements of NS tides across the masses in this sample will constrain their radii to better than 100 m — one part in 100 [107, 108] (Fig. 4, top right). This, in turn, will result in the identification of features in the NS mass-radius relation that can reveal quantum chromodynamics phase transitions, and a population-wide constraint on the NS radius (for the common NS equation of state (EOS)) at the 10 m-level, revolutionizing our knowledge of high-density matter [CESL7, CESL15, 109, 110] (Fig. 2). A network including the two CE observatories will localize many nearby BNSs to better than 1 $\mathrm{deg}^2$ in the sky (Fig. 2 and Fig. 4), linking the properties of compact binary progenitors (masses, spins, and tides) to the properties of host galaxies and the diversity of merger outflows—from neutron-rich outflows contributing heavy element nucleosynthesis (e.g., [84, 87, 92]), to radio-to-X-ray emitting jets (e.g., [91, 93–95]). In some cases, GW observations of an inspiralling system can provide the advance notice required to capture light from the moments closest to merger [111–113] (Fig. 2), such as a fast radio burst (FRB) associated with the ejection of the magnetosphere of a hypermassive NS collapsing to a BH [114].

After a BNS merger, oscillations of the hot, extremely dense remnant can produce "post-merger" GWs [115–118]. This heretofore undetected signal probes a region of the phase diagram of dense matter that is inaccessible to collider experiments or direct electromagnetic observations, and where novel forms of matter such as deconfined quarks may appear [CESL15, 119]. CE 40 km and CE 20 km in kilohertz-focused mode (Fig. 3) will provide accurate measurements of the post-merger GW frequencies for events with post-merger network SNR > 5 [118, 120]; for a fiducial simulated post-merger we expect yearly observation (Fig. 2). This will reveal dense-matter dynamics with finite temperature, rapid rotation and strong magnetic fields; shape theoretical models of fundamental many-body nuclear interactions; and answer questions on the composition of matter at its most extreme [CESL6, CESL13, 121]. Direct GW observations of post-merger remnants will help determine the threshold mass for collapse of a rotationally supported NS, with implications for the NS mass distribution (including the NS maximum mass) [122, 123], predictions of electromagnetic counterparts [124, 125], and supernova engine models [CESL1].

In terms of its impact on the multi-messenger science of compact binary mergers (Fig. 2), CE is synergistic with space missions such as *Fermi* [126, 127] and *Swift* [128], the Nancy Grace Roman Space Telescope [129], and future NASA programs focused on the transient and time-variable universe [CESL4, 130, 131]. From the ground, the Rubin Observatory [132], the Extremely Large Telescopes [133, 134], and the next generation Very Large Array (ngVLA) [135] will provide follow-up capabilities for CE discoveries [CESL4, CESL18, 136–138]. The IceCube-Generation 2 neutrino observatory and CE will help constrain emission models for high-energy neutrinos in nearby BNS mergers [9, 139, 140]. Finally, multi-band observations and synergistic data sets can be formed with the LISA space-based GW detector [141–143].

## 1.3. New Probes of Extreme Astrophysics

NSs and BHs, in isolation or in binary systems, can be sources of GW signals that are very different from the signals already detected by LIGO and Virgo [CESL11]. CE, especially in combination with observatories of other messengers, has the potential to use these yet undetected signals to reveal the physics behind a suite of extreme astrophysical phenomena [144, 145]. Here we summarize several predicted "novel" signals, keeping in mind that this collection is likely incomplete.

Spinning NSs produce quasi-periodic GWs that can last for millions of years [CESL11, 144–146]. These signals arise from mass quadrupoles supported by elastic or magnetic stresses, or mass current quadrupoles from long lived "r-modes" (rotation-dominated modes). Accreting NSs are especially driven to nonaxisymmetry by temperature gradients, magnetic bottling, and perhaps r-modes [145]. Extrapolating from the sensitivities of current searches (e.g., [147]), CE is likely to detect multiple accreting NSs under the assumption that their spins are regulated by GW emission [148–150]. Accreting NSs are believed to become "millisecond pulsars" after accretion ends. There is tantalizing evidence that they spin down at a minimum rate consistent with GW emission from a quadrupole sustained by stresses due to a young pulsar's magnetic field mostly buried under the accreted material [151]. With such fields, CE should detect dozens of known millisecond pulsars (Fig. 2) [151], with the potential to detect many more since the upcoming Square Kilometre Array and ngVLA are predicted to discover several pulsars for every one currently known [136, 152]. Non-detection of accreting NSs or millisecond pulsars would strongly confront current theories of their evolution. All-sky surveys for yet-unknown NSs may yield more than a hundred detections with CE [153], with arcsecond localization to guide follow-up searches for pulsars in radio and other electromagnetic wavebands. Long-lived GW signals, particularly in tandem with electromagnetic observations, can provide information not only on the NS EOS (at low temperatures inaccessible to colliders), but also NS composition, spin evolution, internal magnetic fields and microphysics (i.e., viscosity, thermal conductivity, and elastic properties of the crust) [145, 146, 154].

Core-collapse supernovae generate bursts of GWs from the dynamics of hot, high-density matter in their central regions [CESL11]. CE will be sensitive to supernovae within the Milky Way and its satellites [155], with an expected rate of one over the planned 50-year lifetime of CE. A core collapse supernova detected by CE would have nearly an order of magnitude higher SNR than with $A^\sharp$ [144], allowing improved waveform reconstruction and characterization of source properties [156]. The detection of a core-collapse event in GWs would provide a unique channel for observing the explosion's central engine and the EOS of the newly formed "protoneutron star", allowing measurement of the progenitor core's rotational energy and frequency measurements for oscillations driven by fallback onto the protoneutron star. A nearby supernova could provide a coincident neutrino detection, giving a spectacular multi-messenger event [144]. Some extreme supernovae, such as collapsars or with "cocoons", could generate GWs that could come into reach with A$^\sharp$ and be probed statistically (at a rate of ~ 10 per year) with CE [CESL12, 157, 158].

GWs are also generated by other dynamic NS events [145] such as magnetar gamma-ray flares (possibly accompanied by FRB) and pulsar glitches. Such impulsive, energetic events will excite the many normal modes of NSs, including the strongly GW-emitting "f-modes" (fundamental acoustic modes). Pulsar glitches may also be followed by weeks-long signals as crust and core readjust. Aided by the time and location of the electromagnetic trigger, upgrades to existing observatories can detect GW signals only in the most optimistic scenarios [159, 160], while with CE detection is likely in a wider range of scenarios [CESL2]. Detection of f-modes will measure the cold NS EOS and masses of a population different from that seen in binary mergers, and combined with X-ray observations will yield information on internal magnetic fields [161], while a long post-glitch detection would add information on the viscosity of NS matter. For all burst signals, it is crucial to have multiple GW observatories to provide confidence through coincident detection.

## 1.4. Fundamental Physics and Precision Cosmology

Thanks to its order of magnitude advance in sensitivity over current-generation GW observatories, CE will reveal the physics of strong-field gravity in unprecedented detail via two crucial pathways. First, in three years of operation, a network including a 40 km CE observatory will detect $\gtrsim$ 10 BBHs with a SNR greater than 1000 (the loudest such signal to date, GW200129_065458, had a SNR of 26.8) [7, 162], and hundreds of BBH events with post-inspiral SNR greater than 100 (Fig. 4). Second, CE will detect GWs from sources too rare for us to observe today. In each year of its operation, CE will observe approximately 100,000 BBHs — 1,000 times the total number of GW events observed to date from any source type [163, 164] — providing far more opportunities to discover rare and interesting events. High SNRs and a large number of events will result in roughly two orders of magnitude better theory-agnostic tests of General Relativity (GR) compared to existing facilities [163], possibly revealing physics beyond GR [CESL5, CESL16, CESL17, 163, 165, 166] that is not be accessible to current observatories (§3).

The GW memory effect, a permanent change in strain predicted by GR, is a prominent example of a GR effect that CE will be sufficiently sensitive to detect [CESL5, 165]. As another example, GR mandates that GWs propagate at the speed of light; in the language of quantum field theory that means that the graviton, which mediates the gravitational interaction, must be massless. Owing to the much larger distances it can probe, CE will improve current constraints on the graviton mass by three orders of magnitude [163]. Finally, while GR predicts only two GW polarizations, more general theories of gravity allow for up to four additional vector and scalar modes [166]. The two CE observatories within an international network will be able to discover or improve the limits on the existence of extra polarizations that can be set with advanced detectors [167–169]. The inclusion of a 40 km CE in this network will uniquely differentiate between the two scalar modes [CESL17].

CE will provide a novel and precise measurement of the cosmic expansion rate across the history of the universe [170] with the potential to address the tension in the local cosmic expansion rate inferred by various experiments. Although GW observations of BBHs can be used to measure their luminosity distance, in the absence of additional information the redshift of the source must be inferred from cosmology. However, the combination of a luminosity distance measurement with an independent measure of the source's redshift (either from an electromagnetic counterpart [97] or with other approaches [98, 171–175]) can be used to probe cosmic expansion in ways that are independent of conventional measurements, such as using standard candles and the other elements of the cosmic distance ladder. The many thousands of BNS mergers detected every year by a network containing two CE observatories will have distance uncertainties less than 10% (Fig. 4, lower left). Therefore, CE is well-positioned to improve our understanding of the tension in the local cosmic expansion rate as well as the dark energy equation of state [CESL14, 175–178] and to provide an independent measurement of baryon acoustic oscillations [179]. A network containing the two CE observatories will allow for precise localization of the binary mergers (Fig. 4, top right) and achieve sub-1% precision on $H_0$ in under a year (Fig. 2).

One in a (few) thousand BNS and BBH events can be strongly lensed [180, 181], leading to 50–100 lensed detections in CE annually (Fig. 4, bottom right) [182]. These detections and their multi-messenger counterparts have the potential of unlocking sub-arc-second BBH localization [183] and are unique probes of fundamental properties of GWs and cosmography [184, 185].

We finally note that this aspect of the CE science is synergistic with the goals of the next-generation cosmic microwave background experiment CMB-S4 [9]. Indeed, as discussed in [186], studies of the primordial GW background across a broad frequency range enabled by combining experiments such as CMB-S4 and CE could better constrain cosmological parameters, and particularly the inflationary spectral index and the tensor-to-scalar ratio (Fig. 4).

## 1.5. Dark Matter and the Early Universe

GWs are an exciting new astrophysical probe of dark matter that is complementary to searches at high-energy colliders and underground direct-detection experiments and might reveal the nature of dark matter in several different scenarios [187, 188]. For instance, because of their strong gravitational fields and extreme densities, NSs might capture ambient dark matter over time through scattering off nucleons, or even produce dark matter thanks to the exceptionally high energies achieved in BNS mergers. If a NS were to contain dark matter, it would affect the NS's tidal deformability. The dark matter concentration would likely depend on the NS's age, mass, and environment in this scenario, leading to otherwise inexplicable variations in the tidal deformability [189, 190], and the collapse of NSs to BHs due to dark matter in their cores [191, 192]. These variations will be accessible to CE thanks to the many thousands of high-SNR BNS detections for which tidal deformability can be precisely inferred [164, 193].

BH superradiance is another possible mechanism by which dark matter might generate a GW signature [194–197]. Critically, this mechanism only assumes a coupling through gravity, and as such would still be viable even if dark matter does not have any non-gravitational interaction with baryonic matter. An ultra-light boson with mass in the range $\sim 10^{-13}$ to $10^{-12}$ eV would create a macroscopic "cloud" bound to a BH, which reduces the mass and spin of the host BH. The spin distribution of merging BHs can reveal, or rule out, the existence of these ultralight bosons [198, 199]. The large number of BBHs with good spin measurement (§1.1) implies even a single CE observatory could make a detection—or obtain useful upper limits—in less than a year ([50] and §3.2). In addition, the cloud itself carries a large time-dependent energy density and sources nearly-continuous GWs [200]. These signals can be observed with CE as follow-up searches to rapidly rotating BHs formed in a merger, or with blind searches for continuous waves from sources in the local group [201] (Fig. 4, bottom left) or stochastic waves from nearby BHs [202–206].

CE will also provide a unique opportunity to probe the early universe [207–210]. Standard slow-roll inflationary models are expected to produce a stochastic background with dimensionless energy density $\Omega_{GW} \sim 10^{-17}$ [211, 212], which is too weak to be directly detected by ground-based GW detectors. However, nonstandard inflationary and cosmological models suggest possible backgrounds due to processes such as preheating [213–215], first-order phase transitions [216–220], PBH-seeding multifield inflation [221–223], and cosmic strings [224–228]—all with energy densities within the reach of CE (Fig. 2 and bottom right of Fig. 4). The detection of a cosmological stochastic background would be of fundamental importance for our understanding of the early universe; and even a non-detection would allow for constraints on beyond-standard-model physics at energies orders of magnitude beyond than those accessible with particle accelerators. CE in a network of detectors with comparable sensitivity is key to this goal, as the foreground of resolvable sources must be precisely modeled to reveal the much fainter background [208, 210, 229–232].

# 2. The Cosmic Explorer Concept

The CE concept presented in the CEHS consists of two widely-separated L-shaped facilities in the US, each housing one detector. This pair of detectors maximizes the scientific output with a 40 km arm length detector that is unmatched for deep, broadband sensitivity, partnered with a second detector (20 km) to allow for source localization and polarization sensing, and to provide the capability of tuning its sensitivity to the physics of NSs after they have merged (see §§ 1 and 3) [233]. This concept also takes advantage of efficiencies associated with simultaneous construction, commissioning and operation of two sites within the US, as done by LIGO.

The heart of each CE detector is a dual-recycled Fabry–Perot Michelson interferometer, operated with suspended test masses at room temperature, probed with a 1 μm-wavelength laser, and quantum-enhanced by the injection of frequency-dependent squeezed light. Crucially, this is the same technology used by LIGO to reach unprecedented sensitivity in the O4 observing run. Relying on scaled-up, proven LIGO technology where possible, along with targeted technical advances, provides a straightforward approach to significant improvement with relatively low risk (see also §4.5).

## 2.1. Sensitivity

The expected GW strain sensitivity of CE is shown in Fig. 3, together with an estimate of the ultimate performance of LIGO A+ and A$^\sharp$. The 40 km CE detector in its default broadband configuration reaches a strain sensitivity of about $2.5 \times 10^{-25} / \sqrt{\mathrm{Hz}}$ over a wide band, providing an order of magnitude improvement over A+ at 100 Hz, increasing to 50 times at 20 Hz. Three key factors deliver the superior sensitivity and increased bandwidth of CE:

**Order-of-magnitude longer arms —** Detector arms up to ten times longer than LIGO's boost the sensitivity with relatively low risk, while at the same time matching the GW antenna size to the shortest expected signal wavelength [234]. As a result, one of the two dominant noise sources in A+, thermal noise in the test mass optical coatings, is reduced twenty-fold due to the larger optical beams and the longer arms, while most other displacement noise couplings are suppressed ten-fold [1, 235].

**Quantum sensing —** The other dominant noise source over much of the observation band comes from quantum vacuum fluctuations of the optical field. While also reduced by the longer arms, this quantum sensing noise can be further reduced by quantum vacuum squeezing. The recent success of frequency-dependent squeezing technology [236], with a 2× quantum noise reduction at the LIGO observatories, suggests that the 3× quantum noise reduction assumed for CE is within reach.

**Improved low-frequency isolation —** Targeted isolation system improvements further reduce the noise at and below 20 Hz beyond the order-of-magnitude reduction that comes from arm length alone [237]. The test masses will be isolated from seismic disturbances with both passive and active systems scaled up from those in Advanced LIGO [238, 239], and equipped with improved sensors [15, 240]. A dedicated seismometer array will be used to measure the local seismic field, enabling the subtraction of noise introduced via direct gravitational coupling of ground motion to the test mass ("Newtonian" or "gravity gradient" noise) [241, 242]. Finally, longer and heavier multiple pendulum suspensions will suppress environmental vibrations and the suspensions' thermal noise.

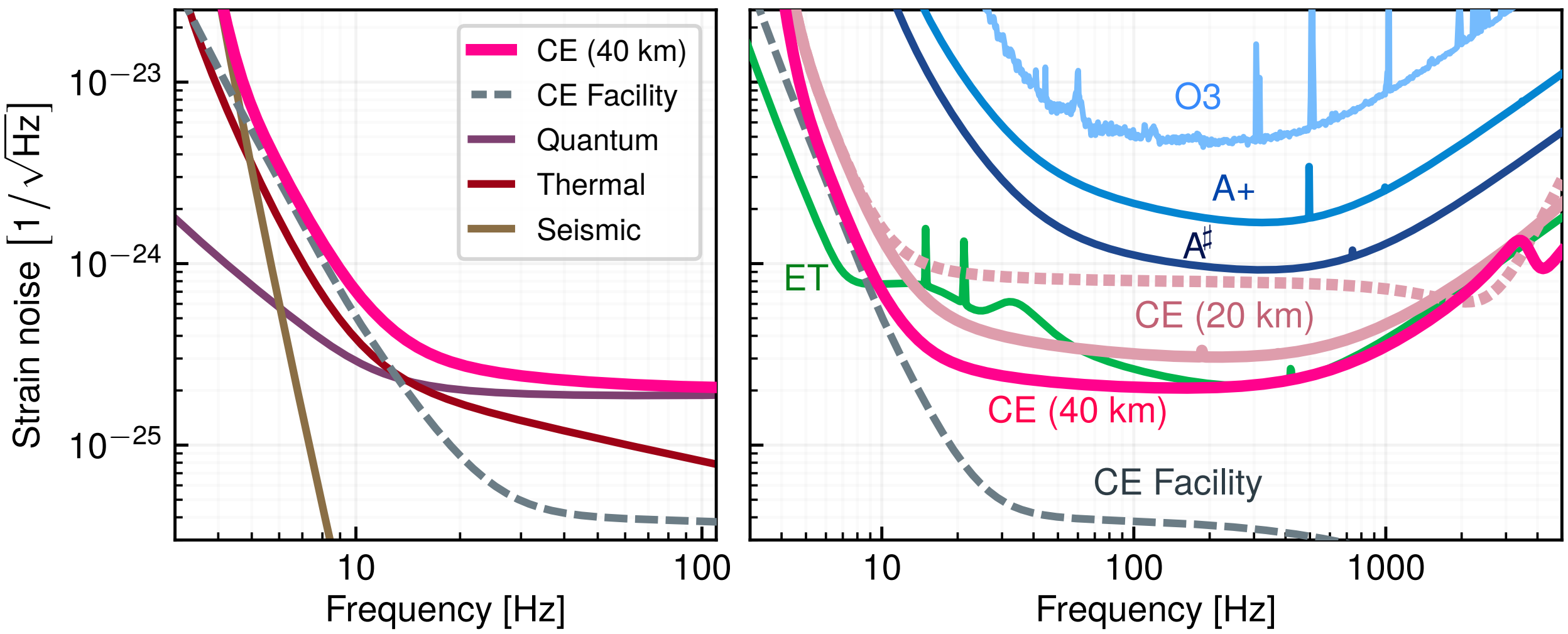

**Figure 3:** *Left:* The spectral sensitivity of CE and the known fundamental sources that contribute to the total noise. *Right:* Measured sensitivity of LIGO in its third observing run (O3) and estimated sensitivities of LIGO A+, LIGO A$^\sharp$, Einstein Telescope (ET) (assuming that the six independent interferometers that form ET are all operating), and the 20 km and 40 km CE detectors. By reconfiguring several smaller optics, the 20 km detector could be operated either in a broadband mode (solid) or a kilohertz-focused mode (dotted). The 40 km facility limit is indicated with dashes.

## 2.2. Technology

The CE test masses will be significantly larger and heavier than in LIGO A+ (see Table 1) — reducing coating thermal noise through larger laser spot size and displacement noises through greater inertia — requiring a focused development effort for manufacturing, polishing, and coating the larger optics. These larger optics will be suspended and seismically isolated to lower frequencies, requiring larger suspensions and seismic isolation platforms with an increased payload capacity. To reduce the quantum sensing noise, high circulating arm power (1.5 MW, a four-fold increase with respect to the maximum power achieved in current detectors) and high squeezing levels (10 dB, see Table 1) are required to meet CE sensitivity targets. Advancements in control strategies will be necessary to stably and reliably operate at such high power and squeezing levels — in particular, thermal and radiation pressure effects on the optics will have to be managed. Finally, with the longer arms comes a greater infrastructure cost. While the vacuum design is informed by the LIGO experience [1, 243], R&D is underway[2] to reduce cost through value engineering.

**LIGO A$^\sharp$ as a CE Pathfinder —** Most of these CE technologies can be at least partially demonstrated within the limits imposed by the LIGO facilities. This idea grew into the envisioned LIGO A$^\sharp$ upgrade [15, 16], which not only boosts the scientific output of the current LIGO facilities, but also acts as technology pathfinder for CE. Table 1 lists the key design parameters of LIGO A+, LIGO A$^\sharp$ and CE, highlighting A$^\sharp$ as a stepping stone towards CE. A full description of enabling technologies is present in the CEHS [1], and in the living document [244].

[2]NSF Award PHY–2207475, *Enabling Research for the Third Generation Gravitational Wave Detectors*, PI: Lazzarini; Co-PI: Weiss.

**Table 1:** The main detector design parameters for A+, A$^\sharp$, and CE. Common to all are 1 µm laser wavelength, and fused silica test masses operated at room temperature. "Test mass coatings" refers to the thermal noise level of the test masses; "A+" thermal noise is a factor of 2 lower than (current) Advanced LIGO, and "A+/2" is another factor of 2 below that. The Newtonian noise mitigation is given for Rayleigh waves, and includes both passive and active measures.

| Design parameter | A+ | A$^\sharp$ | CE |
|---|---|---|---|
| Arm length | 4 km | 4 km | 20 km, 40 km |
| Arm power | 750 kW | 1.5 MW | 1.5 MW |
| Squeezing level | 6 dB | 10 dB | 10 dB |
| Test mass mass | 40 kg | 100 kg | 320 kg |
| Test mass coatings | A+ | A+/2 | A+ |
| Suspension length | 1.6 m | 1.6 m | 4 m |
| Newtonian mitigation | 0 dB | 6 dB | 20 dB |

### 2.3. Facility Limits

The CE facilities — that is, the L-shaped civil infrastructure, including large vacuum system and associated experimental chambers — will constitute a major investment and are expected to have a 50-year lifetime.[3] With this in mind, they will be designed to be flexible enough to support advancements in detector technology during this period. Two potential near-term upgrades are alternative coating materials, such as crystalline GaAs/AlGaAs [245], that could provide much lower coating thermal noise (especially relevant for the 20 km detector), and a combination of higher laser power and lower optical losses with high-fidelity squeezed states to reduce the quantum noise. Longer-term upgrades might include cryogenics or alternate optical configurations [246, 247]. Figures 1 and 3 highlight the Facility Limit, i.e., the sum of infrastructure-specific noise sources that would be common to all future detectors utilizing the CE infrastructure, indicating that these facilities could support an additional factor of five improvement in sensitivity relative to Fig. 3.

## 3. Impact of Network Configurations on Science Goals

The scientific potential of CE is vast, and the science that can be anticipated is groundbreaking on many fronts (§1). Essential to fulfilling the majority of CE's science objectives is the ability to localize sources in the sky, and to measure their properties, such as distances, redshifts and masses. If two or more GW observatories detect a compact binary, its sky location and orientation relative to the line of sight can be inferred from arrival time delays at different detectors, from signal strength consistency with the antenna patterns, and through observation of both GW polarizations. This significantly improves sky localizations and distance measurements, which in turn are needed to estimate the source-frame masses. This section summarizes the scientific potential of a range of global GW network configurations for which we have performed a dedicated trade-study that is intended to directly addresses the charge of the NSF MPSAC ngGW subcommittee [248].

[3] For comparison, the 1990s LIGO vacuum systems are expected to last beyond 2040 with continued maintenance.

**Table 2:** We consider four classes of networks containing zero to three XG observatories. The HLA (Hanford–Livingston–Aundha) network represents existing or under-construction LIGO observatories that may operate at A$^\sharp$ sensitivity in the XG era, and sets a baseline for assessing CE's return on investment. 40LA and 20LA represent a single CE observatory operating with an A$^\sharp$ network. 4020A is the CE reference configuration, operating with an upgraded LIGO Aundha in India (LAO), while 40LET and 20LET represent a single CE observatory operating with LLO at A$^\sharp$ sensitivity and ET. 4020ET is the reference CE configuration operating with ET.

| Num. XGs | Name | Detectors in the network |
|---|---|---|
| 0 | HLA | LHO, LLO, LAO |
| 1 | 20LA | CE A 20 km, LLO, LAO |
| | 40LA | CE A 40 km, LLO, LAO |
| 2 | 20LET | CE A 20 km, LLO, ET |
| | 40LET | CE A 40 km, LLO, ET |
| | 4020A | CE A 40 km, CE B 20 km, LAO |
| 3 | 4020ET | CE A 40 km, CE B 20 km, ET |

## 3.1. Gravitational-Wave Observatory Network Configurations

The ngGW charge asks the subcommittee to consider Next Generation (XG) (at least 10× the sensitivity of LIGO A+) US observatories as part of an international network as well as potential upgrades to the current LIGO detectors (such as, A$^\sharp$). Our models for each of these network nodes are described below and summarized in Table 2. We note that the critical feature of a future network is the number of GW detectors present. Their locations are of secondary importance [249].

**Cosmic Explorer Observatories (CE A, CE B) —** Since the locations of the CE observatories have yet to be determined, we selected two fiducial locations for CE; CE A off the coast of Washington state, and CE B off the coast of Texas. These locations are intentionally unphysical to avoid impacting our ability to find a home for CE (§4.3), but close enough to a wide range of potential sites to be representative from the point of view of GW science. The CE A location is considered in both the 40 km and the 20 km lengths, while the CE B location hosts only a 20 km observatory.

**Existing LIGO Sites (LHO, LLO, LAO) —** In order to focus on the science enabled by CE beyond what is possible in the current facilities, we model the LIGO detectors in an upgraded form (known as "A$^\sharp$", which has comparable sensitivity to the cryogenic "Voyager" configuration [15]) that approximately represents the limit to what is achievable in the LIGO facilities. Furthermore, in addition to the LIGO Hanford (LHO) and LIGO Livingston (LLO) detectors, we also consider LIGO Aundha (LAO) in the A$^\sharp$ configuration, as it is expected be operational in the early 2030s.

**Einstein Telescope (ET) —** ET is a planned XG GW observatory in Europe [250]. It is currently envisioned as an underground triangular facility with 10 km arm length, housing six interferometers. The targeted timeline calls for first observations by the mid-2030s. The underground location, which is strongly preferred in Europe, also suppresses the expected seismic disturbances, thereby reducing the Newtonian noise that limits ground-based GW facilities at low frequencies (cf. the difference between CE and ET below 8 Hz in Fig. 3). While we are encouraged by ET's adoption into the European Strategy Forum on Research Infrastructure (ESFRI) road map, we present some network configurations that do not include ET to highlight the value of US investment even in the absence of our European collaborators. LIGO's two US-based observatories with common management has proven to be a very successful model we wish to emulate.

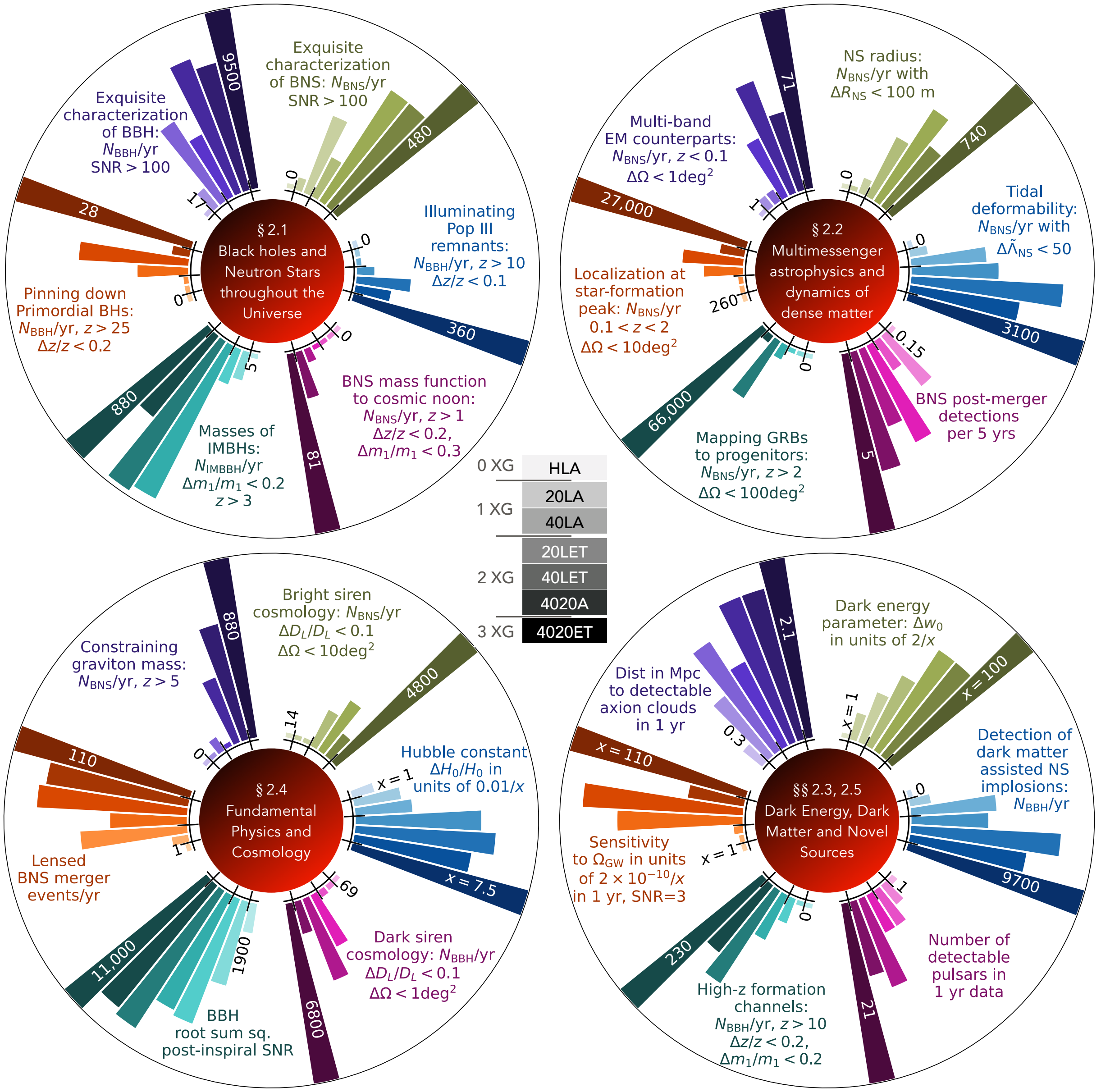

**Figure 4:** Polar histograms (linear scale) showing how CE can accomplish the key science goals discussed in §1. Observatory networks are as in Table 2. Broadly, these histograms show that CE in its reference design, operating with at least one 4 km LIGO A$^\#$ detector, will achieve the key science goals in §1, and that its scientific output would be further enhanced as part of an international network. *Top Left:* An XG network is critical to making high-fidelity observations (SNR > 100) of BH and NS populations, including primordial and Population III BBHs, while accurately measuring their masses, redshifts, and locations in the sky. *Top Right:* The HLA network cannot facilitate the electromagnetic follow-up of mergers at the highest redshifts accessible to the best telescopes while an XG network will routinely provide alerts to such mergers. XG observatories can make exquisite measurements of the radius of NSs and their tidal deformability, and detect post-merger signals from merger remnants. *Bottom left:* Precision tests of GR are enabled by extremely high-fidelity events (SNR > 1000), and also by combining data from thousands of lower-SNR events, producing root-sum-square SNR > 10 000 in the post-inspiral phase of BBHs. Additionally, thousands of BNS and BBH detections with accurate measurements of the distance and sky-localization facilitate precision cosmology and a few hundred strongly lensed events would provide fundamental probes of GWs and cosmography. *Bottom right:* The XG network has abundant discovery potential with the ability to measure the dark energy equation of state parameter $w_0$ (and its variation with redshift [164]), observe weak and rare signals (e.g., pulsars), speculative sources (e.g., BNSs converted to BBHs due to accumulation of dark matter), primordial backgrounds and an opportunity to discover physics beyond the Standard Model (e.g. axion clouds).

## 3.2. Impact on Science Goals

The relative performance of different detector networks was assessed with the Fisher matrix approach using the open-source GWBENCH software [53, 251]. Results reported here are broadly (within 20%) consistent with those found by other authors using the same approach [252–254]. In Fig. 4 we show the relative performance of networks with zero, one, two or three XG observatories (Table 2) with regard to the key science goals described in detail in §1. The various symbols that appear in that figure are as follows: $N_{\rm BNS}$, $N_{\rm BBH}$ and $N_{\rm IMBH}$ are the number of BNS, BBH, and IMBH mergers, respectively; $\Delta$ followed by a symbol refers to the $1\sigma$ uncertainty in the quantity that follows it found using the Fisher matrix approach (except the sky-position uncertainty $\Delta\Omega$, for which it is the 90% credible interval); $\Omega$, $D_{\rm L}$ and $z$ are the source's angular position in the sky, luminosity distance and redshift, respectively; $m_1$ is the mass of the primary companion of a binary; $R_{\rm NS}$ and $\tilde{\Lambda}_{\rm NS}$ are the radius and dimensionless tidal deformability of a NS, respectively; $\Omega_{\rm GW}$ denotes the energy density in stochastic GW background relative to the closure density of the universe; and $w_0$ is the dark energy equation of state parameter.[4]

In all cases, we take a network of three $A^\sharp$ detectors with no XG observatories as our baseline and show the improvements in various science outcomes obtained from networks containing one or more CE observatories. The addition of CE observatories to the global network provides a significant enhancement in their ability to achieve the science targets detailed in §1. At a minimum, we obtain a factor of ~10 improvement in the various metrics and, in many cases, XG detectors facilitate observations that are simply not possible with the $A^\sharp$ network, as shown in Figs. 2 and 4.

A network of $A^\sharp$ detectors with no XG observatory provides moderate gains over the A+ network, allowing, e.g., observation and localization of BNS mergers to redshift $z \approx 0.3$ and BBH mergers to $z \approx 2$. A single XG observatory greatly extends the reach of the network, with BNS mergers observable to the star formation peak at $z \approx 2$ and BBHs observable to $z \gtrsim 10$, the epoch of the first stars in the universe (Fig. 1 and §1.1). This will vastly increase the rate of signals, as well as enable observations of nearby events with unprecedented fidelity. Consequently, for science goals which require the observation of new signals, such as continuous GWs from pulsars (§1.3), or of new features in signals, such as the BNS post-merger signal (§1.2), a single XG observatory is transformational. A single CE20 in kilohertz-focused mode (Fig. 3) will observe one post-merger event every few years with a network SNR> 5. The expected observing time required to achieve these science goals is reduced by at least an order of magnitude with a single CE observatory complementing the A$^\sharp$ network. Similarly, the number of BNS and BBH mergers observed at SNR > 100 will increase by an order of magnitude, enabling precision measurements of NS radii (§1.2) and comparisons between observations and GR predictions (§1.4).

[4] The technical details of our study are summarized as follows, and will appear in a forthcoming report. We simulate a local population of BBHs with mass, spins and redshift distributions consistent with Ref. [44]. Pop III BBHs are assumed to have a fixed primary mass of $20\,M_\odot$, a mass ratio of 0.9, a spin distribution consistent with what was found in Ref. [44], and a redshift distribution given in Eq. (C15) of Ref. [51]. PBH binaries have masses drawn from the log-normal distribution of Ref. [5], zero spin, and the redshift distribution given by Eq. (5) of Ref. [5]. We use a merger rate for the local population that is consistent with Ref. [44], resulting in 96 000 sources per year. Following Ref. [5], we obtain 2400 and 600 Pop III and PBH mergers per year, respectively. For BNSs we assume a double Gaussian with median values from [255]; spins uniform in $[-0.1, 0.1]$ and aligned with the orbital angular momentum; the APR4 nuclear EOS and the same redshift distribution as the local BBH population. This gives 1.2 million BNSs per year. Neutron star – black hole binaries (NSBHs) have BH masses drawn from the POWERLAW+PEAK distribution from Ref. [44], NS masses are uniform in the range $[1, 2.2]\ M_\odot$; BH spins aligned with the orbit and drawn from a Gaussian centered at 0, with standard deviation of 0.2, NS spins aligned with the orbit and in the range $[-0.1, 0.1]$ and the same redshift distribution as the local BBH population. We use a local NSBH merger rate of $45\,{\rm Gpc}^{-3}\,{\rm yr}^{-1}$ [44], which gives 180 000 sources per year. Finally, we simulate a population of IMBHs with masses drawn from a power law with index $-2.5$ in the range $[100, 1000]\ M_\odot$; spins uniform in the range $[-0.9, 0.9]$ and redshifts distributed as the local BBH population. For detection of a compact binary signal we require a minimum SNR=5 in each observatory and a network SNR=10. For a post-merger signal, we require a minimum network SNR=5 and we assume a kilohertz-focused mode for CE20 (pink-dashed line in Fig. 3). For pulsars we assume one year of observation and a quadrupole ellipticity of $10^{-9}$ [151], with spin frequencies taken from the ATNF catalog [256].

Several science goals require the accurate localization of binary sources (both in the sky and in distance/redshift) to infer their intrinsic masses. For example, a source at redshift $z = 10$ observed in a single XG observatory would be essentially unlocalized in the sky, and have a distance uncertainty of ∼ 50%. This leads to an uncertainty of ±4 in redshift measurement and, due to the mass–redshift degeneracy in GW observations, a 40% uncertainty in the mass measurement, rendering a detailed study of the BBH population at high redshifts impossible. For events that lie beyond the A$^\sharp$ horizon, accurate localization and mass/redshift measurements can only be achieved with a network of two or more XG observatories. A second XG observatory enables at least partial localization of large numbers of events and provides a substantial improvement in the GW measurement of the Hubble constant and other cosmological parameters (§1.5). A two XG network also enables precision localization of nearby BNS. A network of three XG detectors enables good localization of the majority of sources, with tens of thousands of BNS signals each year localized to better than 10 deg$^2$, thereby enabling multi-messenger follow-up observations (§1.2 and Fig. 4). A three XG network can also improve constraints on the binary inclination angle relative to the line of sight, and hence the luminosity distance measurement (§1.1 and §1.4). Multiple XG observatories also improve the confidence in detection of poorly modeled sources (§1.3), enable polarization measurements that are relevant for tests of GR (§1.4), and the inference of the presence of dark matter in NS cores or the detection of primordial stochastic backgrounds (§1.5).

# 4. The Cosmic Explorer Project

While the primary objective of CE is to answer deep scientific questions in fundamental physics, nuclear physics and astrophysics, an undertaking of this scale has impacts well outside of the scientific community. If CE is to be funded by US taxpayers, it must serve the needs of the nation in a broad sense, be cognizant of and responsive to potentially impacted communities, and be designed to maximize the return on taxpayer dollars. This section broadens the view of CE relative to the scientific and technical highlights presented in §1 and §2 with information about the CE Project, cost estimates and timeline (§§ 4.1 and 4.2), the process of finding potential homes for CE (§4.3), education and equity efforts (§4.4), and known project risks (§4.5).

The CE Project was organized in 2021 following the completion of the CEHS [257, 258], and currently has over 40 members with a wide range of expertise. In addition, CE has international partners (Australia, Canada, Germany, UK) and a broad community of over 500 members in the CE Consortium. In Fall 2022, the Project organized the development of seven proposals to the National Science Foundation (NSF) to fund the first three years of the CE conceptual design, six of which have been recommended for funding. The Project also organized the writing of this white paper.

## 4.1. Cost Estimates

CE observatories are envisioned as largely above-ground, L-shaped facilities [1]. This choice is in line with currently-operating observatories, but different from KAGRA in Japan and the planned

Einstein Telescope (ET) in Europe. In the US context, where large relatively flat areas with low population densities can be found (§4.3), building above ground maximizes scientific return on investment by avoiding tunneling costs and the complexity of underground construction, installation and operation.

**Design and Construction costs —** The initial cost estimate for the CE reference concept consisting of a 40 km observatory and a 20 km observatory is approximately $1.6B (2021 USD), as published in the CEHS. The CEHS also presents estimates for two 20 km observatories ($1.3B), a single 40 km observatory ($1.0B) and a single 20 km observatory ($0.7B). These estimates are based on extrapolating actual costs from LIGO construction, the Advanced LIGO upgrade, and the work of professional civil engineering and metallurgy consultants. Cost drivers for a CE observatory are arm length, beamtube material and diameter, and location-dependent civil-engineering costs. Many of the costs associated with arm length are proportional to the length (e.g., the beamtube and its enclosure, the roads along the beamline, electrical utilities along the beamline, the slab supporting the beamtube), and largely location independent (within 10% of the national average). The cost of excavation and transportation is generally not proportional to the length of the facility, and highly dependent on topography and geology (e.g., depth to rock). Notably, the cost of the detectors is not a leading driver (estimated at ≈ 28% of the total; see [1] for a full breakdown of this cost estimate).

**Maintenance and Operations Costs —** Again drawing from the CEHS, yearly maintenance and operations costs for the CE reference concept (2 observatories) were estimated at $60M (2021 USD) [1]. This estimate is based on LIGO experience[5] and includes the observatory facilities, vacuum systems, and detector hardware. It also includes management, community engagement, and the data analysis and curation required to make CE data available and accessible to the scientific community and the public. Notably, this estimate does *not* include university research or development efforts towards future CE upgrades, and it assumes a model in which much of the data analysis (beyond that needed to issue astronomical alerts) happens in the scientific community (and is separately funded). To respond directly to the ngGW subcommittee's charge, we compute "maintenance and operations costs for the first ten years" as $670M in 2023 USD. If we assume 3% inflation for all future years and compute operation costs from 2035 to 2045, the total is $1.1B in then-year dollars.

## 4.2. Timeline

The timeline for CE spans multiple decades and takes place in distinct stages. The development stage for CE began in 2013 and culminated in the publication of the CEHS in 2021. The design and site-selection stage is expected to start this year (2023) and continue for 8 to 10 years. Expedient funding will allow CE construction in the early 2030s, and *initial observations in the middle of the next decade*. Figure 5 graphically summarizes this timeline.

In parallel with these technical efforts, work on relationship building with local and Indigenous communities that are engaged with and/or impacted by CE, from local to global, will be of ever increasing importance. The CE project will partner with communities at potential observatory host locations to pave mutually beneficial pathways forward beginning with location identification (§4.3).

## 4.3. Site Evaluation and Community Partnerships

Construction on the scale of the CE is not only technically challenging, but requires attention to potential social, cultural, and economic impacts during all stages, from design to divestment.

[5] LIGO's yearly operating cost will be $50M in 2024 for two 4 km Observatories; this includes R&D for future upgrades, not foreseen for CE, and is not dependent on the specific detector design.

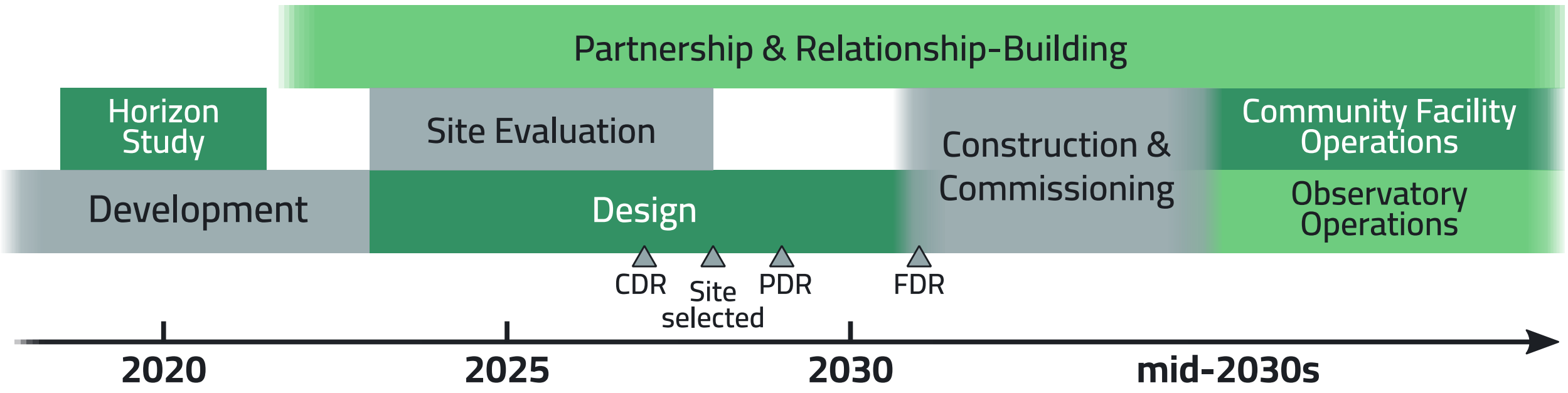

**Figure 5:** CE top-level timeline showing a phased approach to design and construction. Following the NSF Research Infrastructure Guide [259], the design stage has 3 major milestones: conceptual design review (CDR), preliminary design review (PDR, budget is final), final design review (FDR). The timeline shown here assumes an aggressive funding model, with construction in the early 2030s and operation in the mid-2030s. While the initial mandate is expected to be for 20–25 years, the facility may operate for 50+ years [1]. The eventual divestment stage is not indicated.

During the CE development stage, many physically promising locations in the US that could plausibly accommodate a 20 km or 40 km baseline observatory were algorithmically identified using publicly available topological and land use data. Several of these locations were followed up with additional publicly available data, such as land ownership, current and traditional Indigenous connection to land, proximity to cities and seismicity, as well as earthquake, flood and wind hazard. Fundamental physical requirements for a CE observatory include quiet ground motion and other favorable ambient environmental conditions, modest weather disturbances, minimal susceptibility to natural disasters, and low human-induced noise. Optimal locations would also have access to a number of strategic infrastructures — such as roads, rail, airports, and cities — to support all phases of the project, from construction (i.e., delivery of vacuum pipes and large equipment) to commissioning and operation.

Following the successful example of the LIGO Livingston Observatory in Louisiana, CE presents an opportunity to broaden participation and build research competitiveness and STEM capacity in states that have traditionally been awarded less NSF support; a number of plausible locations are in EPSCoR jurisdictions [260]. Furthermore, cost estimates made for the CEHS indicate that excavation will not be the leading cost driver for locations with favorable geology and topography — meaning that there is some flexibility to choose locations that do not have the lowest excavation costs if they excel in other areas, such as social and environmental consideration [1].

Beyond the physical features of a location, decades of LIGO operations have also highlighted the importance of the social context of an observatory. CE has the unique opportunity to prudently reconceptualize community engagement and to respectfully work within the local and global socio-cultural context. In today's social and legal context (e.g., [9] and references therein), an investment in establishing robust and sustained relationships with local and Indigenous communities will be fundamental to the evaluation of each location's potential for housing an observatory.

In parallel with the design stage of CE, a deeper, broader, and more culturally aware study of potential locations in the US will be required. In anticipation of near-term funding, work has begun to assemble a team with diverse expertise in GW experiment and computation, astrophysics, geology, geography, and sociology. Through the CE Director of Community and Land Partnership [257], project leadership has initiated work with consultants in law and economics to develop an integrated, interdisciplinary approach to location evaluation that supports CE's scientific goals while simulta-neously identifying areas of relevance to, and potential synergies with, Indigenous and other local communities. Identifying and evaluating the most promising locations for CE observatories will take place in concert with developing protocols and best practices for large-scale projects to be in partnership with local and Indigenous communities. This work will draw on expertise from traditional knowledge (as appropriate), physics, geology, Geographic Information Systems (GIS), and sociology to incorporate both quantifiable (e.g., topology and seismology) and go/no-go information (e.g., protected lands or historical lands) into the evaluations that identify initial locations of interest. Results from subsequent assessment steps will be used to refine the collection of viable locations toward a set of well characterized candidates in the late 2020s (§4.2).

## 4.4. Broadening Participation and Cultivating a Thriving Community

CE will be a tremendous investment in the US scientific workforce. The CE Project is committed to the equity advancing values emphasized in the NSF 2022–2026 Strategic Plan, including advancing the "missing millions" of underrepresented women and communities of color who diversify the US STEM workforce [261]. Building upon the NSF Strategic Plan's vision and guidelines from the Astro2020 Panel on the State of the Profession and Societal Impacts [9], the CE Directors' Office provides central leadership and integration into CE structure and culture through the Director of Equity, Diversity, and Inclusion (EDI) [257]. Coordination across CE Project institutions includes facilitating training with external experts, developing mentoring structures connected to recruiting, hiring, retention, and promotion for CE members and leadership, and development of the CE code of conduct and ombuds office. Project leadership also facilitates partnerships between institutional Research Experiences for Undergraduates (REU) and Bridge programs to bring research opportunities to undergraduate and graduate students. Opportunities within the CE Project already include the International REU program at University of Florida, programs at the Hispanic and Native American Pacific Islander-serving institution California State University, Fullerton, and research activities at Emerging Research Institutions Syracuse University and Texas Tech University. These efforts serve the need of the nation by contributing to the development of the STEM workforce, and directly address a Project risk associated with the multi-generational nature of CE (§4.5).

The University of Washington Center for Evaluation and Research for STEM Equity (CERSE) [262] collaborates with CE leadership to design and execute project consulting, including collection and analysis of demographic data and advising on organizational development. The CE Project is connected to the broader community through the Gravitational Wave International Committee (GWIC) [263], the Multimessenger Diversity Network (MDN) [264], the Gravitational Wave Early Career Scientists (GWECS) [265], and the GW Allies [266] to facilitate the sharing of resources and best practices. CE community engagement encompasses contributions to STEM workforce development, connections to the broader astrophysics community, and to communities near potential CE host locations (§4.3).

## 4.5. Risks and Mitigation Strategies

A project of the scale and complexity of CE will have a number of risks at each phase of development. The Project will manage these with well-established practices, leveraging experience from LIGO and its upgrades. An initial assessment of risks and mitigations can be found in §11.4 of the CEHS. In general, the CE observatory has been conceived to minimize risk for the infrastructure by evolving from the successful LIGO design, and minimizing risk for the detector by planning on re-scaling the LIGO designs to the greater CE length (§2) with two observatories sharing common design and management teams. The two observatory teams will profit from the same synergy seen in LIGO. With that basis, we discuss some leading risks below.

A leading technical risk for both the 40 and 20 km instruments is in making the larger diameter optics to the required optical performance, although preliminary contacts with vendors indicate that CE optics should only require a modest investment in retooling. There are also technical risks associated with the interferometer control and stability associated with scaling-up to CE (e.g., frequency control bandwidth limits due to arm length, parametric instabilities in large mirrors, etc.). These are known challenges that will be properly addressed by research during the conceptual design phase of the project.

Mitigation of management and non-technical risks is also important to CE success. In particular, site identification and preparation for CE will not only require that technically suitable locations be found, but also the development of enduring relationships with the local and Indigenous communities (§4.3). The significant duration of CE requires that the engaged scientific and engineering team be multi-generational; to ensure this, CE involves a range of teaching institutions distributed across the US, and the team will maintain a vigorous program of research involving students throughout the Project duration and into the observing epoch (§4.4).

Lastly, there is the risk of missing significant added science that is enabled by other GW observatories (most notably ET; §3.1), or complementary photon and particle observatories (§1). We address these external risks by maintaining close relationships with these projects, communicating our plans and capabilities, and helping to demonstrate to funding agencies the synergistic potential of observing with CE as part of a global multi-messenger network.

# 5. Outlook

The CE Project is preparing for entry into the observatory design stage. The conceptual and preliminary design phases (the next 8–10 years) will see the development of detailed instrument, vacuum system and facility designs, as well as accurate cost and schedule estimates. CE's order of magnitude sensitivity improvement over LIGO A+ relies on proven technology and decades of experience with the LIGO observatories. The CE detector will profit from the $A^\sharp$ development, ensuring readiness with limited R&D and low risk. Site identification and evaluation will take place during the design phases, as will economic, environmental and socio-cultural impact studies.

The CE concept (two observatories, one 40 km long and one 20 km long) was developed over the last decade, leveraging broad input from the scientific community [1, 11–14, CESL1–CESL19]. The CE project will continue to solicit and respond to the scientific community's input, in part through the CE Consortium. Establishing partnerships with local and Indigenous communities will be crucial for the project's success.

Once operational, the cosmic reach and exquisite sensitivity of CE will revolutionize our understanding of the universe while continuing the United States' leadership in gravitational-wave science. CE's extraordinary scientific potential will open doors for discovery in the evolution of our Universe, its contents, and governing laws. Throughout its lifetime, CE will invest in a community-based model [267, 268] and broaden participation in cutting-edge scientific research to empower STEM workforce development for decades to come. If CE is robustly funded through the 2020s, CE's first scientific observations could take place in the mid-2030s. At that time, and as part of an international next-generation multi-messenger network of observatories, CE will bring its generational advance in observational capacity to a multitude of fields in physics, astronomy, and cosmology.

# Acronyms

**BBH** binary black hole 4, 5, 8, 9, 14–16
**BH** black hole 4–7, 9, 14, 15
**BNS** binary neutron star 3, 6, 7, 9, 14–16
**CE** Cosmic Explorer 4–20
**CEHS** Cosmic Explorer Horizon Study 4, 10, 12, 16–19
**EOS** equation of state 6–8, 15
**ET** Einstein Telescope 11, 13, 17, 20
**FRB** fast radio burst 6, 8
**GR** General Relativity 8, 14, 16
**GRB** gamma-ray burst 6
**GW** gravitational wave 4, 6–10, 12–16, 18, 20
**GWIC** Gravitational-Wave International Committee 4
**IMBH** intermediate mass black hole binary 5, 15
**JWST** James Webb Space Telescope 5
**LIGO** Laser Interferometer Gravitational Wave Observatory 4, 5, 7, 10–14, 17–20
**MPSAC** Mathematical and Physical Sciences Advisory Committee 12
**ngGW** Next-Generation Gravitational-Wave Observatory 12, 13, 17
**ngVLA** next generation Very Large Array 7
**NS** neutron star 4, 6–10, 14–16
**NSBH** neutron star – black hole binary 15
**NSF** National Science Foundation 12, 16, 18, 19
**PBH** primordial black hole 5, 10, 15
**SNR** signal-to-noise ratio 3, 5–9, 14–16
**US** United States 5, 10, 13, 16–20
**XG** Next Generation 13–16

# Authors

Matthew Evans,[1] Alessandra Corsi,[2] Chaitanya Afle,[3] Alena Ananyeva,[4] K.G. Arun,[5,6] Stefan Ballmer,[3] Ananya Bandopadhyay,[3] Lisa Barsotti,[1] Masha Baryakhtar,[7] Edo Berger,[8] Emanuele Berti,[9] Sylvia Biscoveanu,[1] Ssohrab Borhanian,[10] Floor Broekgaarden,[8] Duncan A. Brown,[3] Craig Cahillane,[3] Lorna Campbell,[4] Hsin-Yu Chen,[11] Kathryne J. Daniel,[12] Arnab Dhani,[13] Jennifer C. Driggers,[14] Anamaria Effler,[15] Robert Eisenstein,[1] Stephen Fairhurst,[16] Jon Feicht,[4] Peter Fritschel,[1] Paul Fulda,[17] Ish Gupta,[6] Evan D. Hall,[1] Giles Hammond,[18] Otto A. Hannuksela,[19] Hannah Hansen,[14] Carl-Johan Haster,[20] Keisi Kacanja,[3] Brittany Kamai,[21,22] Rahul Kashyap,[6] Joey Shapiro Key,[23] Sanika Khadkikar,[6] Antonios Kontos,[24] Kevin Kuns,[1] Michael Landry,[14] Philippe Landry,[25] Brian Lantz,[26] Tjonnie G. F. Li,[27,28] Geoffrey Lovelace,[34] Vuk Mandic,[29] Georgia L. Mansell,[3] Denys Martynov,[30] Lee McCuller,[4] Andrew L. Miller,[31,32] Alexander Harvey Nitz,[3] Benjamin J. Owen,[2] Cristiano Palomba,[33] Jocelyn Read,[34] Hemantakumar Phurailatpam,[19] Sanjay Reddy,[35] Jonathan Richardson,[36] Jameson Rollins,[4] Joseph D. Romano,[2] Bangalore S. Sathyaprakash,[6,16] Robert Schofield,[14] David H. Shoemaker,[1] Daniel Sigg,[14] Divya Singh,[6] Bram Slagmolen,[37] Piper Sledge,[38] Joshua Smith,[34] Marcelle Soares-Santos,[39] Amber Strunk,[14] Ling Sun,[37] David Tanner,[17] Lieke A. C. van Son,[8,40] Salvatore Vitale,[1] Benno Willke,[41] Hiro Yamamoto,[4] and Michael Zucker[4,1]

[1] LIGO Laboratory, Massachusetts Institute of Technology, Cambridge, MA 02139, USA
[2] Department of Physics and Astronomy, Texas Tech University, Lubbock, TX 79409, USA
[3] Department of Physics, Syracuse University, Syracuse, NY 13244, USA
[4] LIGO Laboratory, California Institute of Technology, Pasadena, CA 91125, USA
[5] Chennai Mathematical Institute, Chennai, India
[6] Institute for Gravitation and the Cosmos, Department of Physics, Pennsylvania State University, University Park, PA 16802, USA
[7] Department of Physics, University of Washington, Seattle, WA 98195, USA
[8] Center for Astrophysics, Harvard & Smithsonian, Cambridge, MA 02138, USA
[9] Department of Physics and Astronomy, Johns Hopkins University, Baltimore, MD 21218, USA
[10] Theoretisch-Physikalisches Institut, Friedrich-Schiller-Universität Jena, Jena 07743, Germany
[11] Department of Physics, The University of Texas at Austin, Austin, TX 78712, USA
[12] Department of Astronomy & Steward Observatory, University of Arizona, Tucson, AZ 85721, USA
[13] Max Planck Institute for Gravitational Physics (Albert Einstein Institute), Potsdam 14476, Germany
[14] LIGO Hanford Observatory, Richland, WA 99352, USA
[15] LIGO Livingston Observatory, Livingston, LA 70754, USA
[16] Gravity Exploration Institute, School of Physics and Astronomy, Cardiff University, Cardiff, CF24 3AA, United Kingdom
[17] Department of Physics, University of Florida, Gainesville, FL 32611, USA
[18] Institute for Gravitational Research, School of Physics and Astronomy, University of Glasgow, Glasgow, G12 8QQ, United Kingdom
[19] Department of Physics, The Chinese University of Hong Kong, Shatin, New Territories, Hong Kong
[20] Department of Physics and Astronomy, University of Nevada Las Vegas, Las Vegas, NV 89154, USA
[21] Department of Astronomy & Astrophysics, University of California Santa Cruz, Santa Cruz, CA 95064, USA
[22] Department of Physics and Astronomy, University of Hawaii at Manoa, Honolulu, HI 96822, USA
[23] University of Washington Bothell, Bothell, WA 98011, USA
[24] Bard College, Annandale-on-Hudson, NY 12504, USA
[25] Canadian Institute for Theoretical Astrophysics, University of Toronto, Toronto, ON M5S 3H8, Canada

[26] Kavli Institute for Particle Astrophysics and Cosmology, Stanford University, Stanford, CA 94305, USA
[27] Institute for Theoretical Physics, Department of Physics and Astronomy, KU Leuven, B-3001 Leuven, Belgium
[28] STADIUS, Department of Electrical Engineering (ESAT), KU Leuven, B-3001 Leuven, Belgium
[29] School of Physics and Astronomy, University of Minnesota, Minneapolis, MN 55455, USA
[30] School of Physics and Astronomy, and Institute for Gravitational Wave Astronomy, University of Birmingham, Birmingham B15 2TT, United Kingdom
[31] Nikhef – National Institute for Subatomic Physics, 1098 XG Amsterdam, The Netherlands
[32] Institute for Gravitational and Subatomic Physics, Utrecht University, 3584 CC Utrecht, The Netherlands
[33] INFN, Sezione di Roma, I-00185 Roma, Italy
[34] Nicholas and Lee Begovich Center for Gravitational Wave Physics and Astronomy, California State University Fullerton, Fullerton CA 92831, USA
[35] Institute for Nuclear Theory, University of Washington, Seattle, WA 98195, USA
[36] Department of Physics & Astronomy, University of California Riverside, Riverside, CA 92521, USA
[37] OzGrav-ANU, Centre for Gravitational Astrophysics, College of Science, The Australian National University, ACT 2601, Australia
[38] Department of Gender & Women's Studies, University of Arizona, Tucson, AZ 85721, USA
[39] Department of Physics, University of Michigan, Ann Arbor, MI 48109, USA
[40] Anton Pannekoek Institute for Astronomy, University of Amsterdam, 1090 GE Amsterdam, The Netherlands
[41] Max Planck Institute for Gravitational Physics (Albert Einstein Institute), Hannover 30167, Germany

Correspondence. ce-questions@cosmicexplorer.org

Cover art. *Rendering of a Cosmic Explorer observatory:* Edward Anaya, Virginia Kitchen, Angela Nguyen, and Joshua R. Smith (CSU Fullerton). *Rendering of gravitational-wave emission from a black hole binary:* scientific contact by Ed Seidel; simulations by Max Planck Institute for Gravitational Physics (Albert-Einstein-AEI); visualization by Werner Benger, Zuse Institute, Berlin (ZIB) and AEI; the computations were performed on NCSA's Itanium Linux Cluster.

# References

## Cosmic Explorer Science Letters

## Other References